\documentclass[11pt]{amsart}

\usepackage{graphicx}
\usepackage{caption}
\usepackage{hyperref}

\DeclareMathOperator{\Pic}{Pic}

\newcommand{\field}[1]{\mathbb{#1}}

\newcommand{\R}{\field{R}}

\newcommand{\set}[1]{\{#1\}}

\newcommand{\beq}{\begin{displaymath}}
\newcommand{\eeq}{\end{displaymath}}
\newcommand{\beqn}{\begin{equation}}
\newcommand{\eeqn}{\end{equation}}
\newcommand{\bPi}{\mathbf{\Pi}}

\newtheorem{theorem}{Theorem}[section]

\theoremstyle{definition}
\newtheorem{defn}[theorem]{Definition}

\newtheorem{prop}[theorem]{Proposition}

\theoremstyle{remark}

\numberwithin{equation}{section}

\title{Geometry, Inference, Complexity, and Democracy}

\author[Jordan S. Ellenberg]{Jordan S. Ellenberg}
\address{University of Wisconsin-Madison}
\curraddr{}
\email{ellenber@math.wisc.edu}
\thanks{}

\subjclass[2010]{Primary }

\date{}

\dedicatory{}

\begin{document}

\begin{abstract}  Decisions about how the population of the United States should be divided into legislative districts have powerful and not fully understood effects on the outcomes of elections.  The problem of understanding what we might mean by ``fair districting" intertwines mathematical, political, and legal reasoning; but only in recent years has the academic mathematical community gotten directly involved in the process.  I'll report on recent progress in this area, how newly developed mathematical tools have affected real political decisions, and what remains to be done.  This survey represents the content of a lecture presented by the author in the Current Events Bulletin session of the Joint Mathematics Meetings in January 2020.
\end{abstract}

\maketitle



\section{What does it mean to be represented?}

Democratic states are founded on the principle that every citizen's views are to be represented in the conduct of the government.  This principle, like most important principles,  is easy to state, difficult to make precise, and almost impossible to implement in a fully satisfying way.

For one thing, democratic states are big.  Even a modestly sized city is large enough that it would be impractical for every decision about zoning, school curriculum, public transport, and taxes to be put to a public plebiscite, let alone to arrive at a consensus.  So modern governments typically operate under some form of representative democracy, in which a small group of legislators are elected by the population to write laws and vote on their passage.  But how to generate this group of popular representatives?  There are a lot of different ways.  In Israel, voters vote for their preferred political party, which is awarded a number of seats in the Knesset roughly in proportion to its share of the popular vote, and then the party chooses the occupants of those seats.  For the Senate of the Phillippines, each voter casts a vote for as many as twelve candidates, and the top twelve vote-getters overall join the Senate.   The most common means of choosing representatives, though, is the one used by the United States Congress and by the legislatures of most of the states; the population is divided up into {\em legislative districts}, and each district chooses a representative by plurality vote.  Under a district system, every voter has a specific representative who, in theory, governs on their behalf and with attention to their interests.

In some systems, this partition of the electorate reflects natural (or at least historically settled) political divisions.  Each U.S. state has two U.S. Senators, because, at least formally, each state is a semi-autonomous political entity with its own particular interests.  (These are thus examples of {\em multi-member districts}, a variant of single-member districting in which each district selects not just one but several representatives in the legislature; multimember districts are also used for a few US state legislatures.)  The partition is almost always along geographic lines, though not always.  In New Zealand, M$\bar{\mbox{a}}$ori people have their own electoral districts, which are superimposed on the general districts; M$\bar{\mbox{a}}$ori voters have the choice in each election whether to vote in the M$\bar{\mbox{a}}$ori or the general district containing their residence.  Or the partition might not have any geographic component at all.  In Hong Kong, there's a seat in the Legislative Council only teachers and school administrators can vote for, one of 30 seats elected by so-called functional constituencies.  The Centuriate Assembly of the Roman Republican had constituencies separated by wealth bracket.  In the upper house of the Oireachtas in Ireland, there is a three-seat constituency consisting of students and graduates of Trinity College Dublin, and another for alumni of the National University of Ireland.

Electoral districts within U.S. states are a different story.  They are patches of land without much meaning.  Nobody in the 2nd Congressional District of Wisconsin, where I live, wears a WI-2 sweatshirt, or could recognize the district from its silhouette.  As for my state legislative district, I had to look it up to be sure I had the number right.  These districts have to be determined somehow, despite lacking robust pre-existing political identities; that is, someone has to select a partition of the population of the state chosen from the ensemble of all possible partitions, a set of unmanageably large size.  This process, historically, has not been the subject of much public attention.  That has now changed.  That's because we now understand something we didn't fully grasp before, which is at least in part a mathematical statement; that the way the population is broken up into districts has an enormous effect on the makeup of a legislature.  

To some extent this is obvious.  If I am in complete control of the districting of Wisconsin, with the power to partition the population any way I wish, and there is a cabal of like-minded people I want to be in control of the state, I could simply make each one of those people their own district, and then create one more district consisting of everybody else.  My hand-picked candidates vote for themselves and then rule the legislature with at most one potential voice of opposition.

That's not fair!  Certainly the people of Wisconsin, with the exception of the cabal itself, would be right to feel themselves unrepresented in the decision-making of the state.

In real life, no one tries to implement a scheme like this.  For one thing, state governments are not allowed to create districts with radically different populations; though until the 1964 Supreme Court decision in {\em Reynolds v. Sims}, state governments in the United States could, and did, do exactly this.   In the United Kingdom, so-called ``rotten boroughs"  with only a few dozen voters were common until the 19th century.

Nowadays, in the United States and many other representative democracies (though not Canada!) districts are fixed by law to be approximately equal in size.  That prevents the kind of cabalization of the legislature I described above.  But it is {\em not}, it turns out, sufficient to keep the choice of partition from having a dramatic influence on the outcome of the election.  The manipulation of district boundaries in order to achieve a desired outcome (most commonly a majority or supermajority of seats for one's own party, or protection of incumbent legislators) is often called {\em gerrymandering}, after a 19th-century Massachusetts governor sometimes thought of as a pioneer of the practice.  In most states, the power to determine legislative districts is held by the legislature itself, creating an obvious incentive and opportunity for a disciplined partisan majority to protect itself from the will of an unfriendly electorate.

We are faced with the following ensemble of questions.  We write $\bPi = \set{\Pi_1, \ldots, \Pi_k}$ for a partition of the state's population into $k$ subsets.  We want to know:

\begin{itemize}
\item What properties should $\bPi$ have in order to be considered ``fair"?
\item Given a proposed $\bPi$, are there quantitative measurements of unfairness which are robust, reliable, and simple enough to be used by judges and courts who have to decide whether $\bPi$ is too unfair to use?
\item In a U.S. context, what kind of constraints on $\bPi$ do the US and state constitutions allow us to impose, and what kind of constraints do those constitutions {\em require} us to impose?  
\end{itemize}

Are these actually math questions?  They are also not {\em not} math questions.  But they have a legal, a political, and a philosophical strand as well, and the strands can't really be unwound from each other.  If mathematicians work on these problems alone, ignoring the other strands, the results are not going to be very useful.  (``Why don't we just draw a grid over the state and make each box a district....?")  But when lawyers and politicians think about redistricting while neglecting the mathematical strand, the result of their work will be no better; and that, by and large, is exactly how these issues have been addressed through most of American history.   In recent years, I am happy to report, there has been a flowering of truly interdisciplinary work, involving both serious mathematics and conscientious attention to political and legal realities, and we have begun to move toward a way of thinking about legislative districting which is sound from all the relevant points of view.  In these notes I'll try to give a brief summary of recent progress in this area.

\section{Measures of fairness}

What do we mean when we say a districting is ``fair" to the residents of a state?  A good way to get a sense of the difficulties here is to contemplate a toy example.  Imagine a state with a population of just 100 people, of whom 60 are members of the Purple Party and 40 vote for the Orange Party.  The population of this state is partitioned into five legislative districts.  Here are four ways the task could be done:

\beq
\bPi_1: \begin{tabular}{c c}
Purple & Orange \\
15 & 5 \\
15 & 5 \\
15 & 5 \\
7 & 13 \\
8 & 12
\end{tabular}
\eeq

\beq
\bPi_2: 
\begin{tabular}{c c}
Purple & Orange \\
9 & 11 \\
9 & 11 \\
9 & 11 \\
17 & 3 \\
16 & 4
\end{tabular}
\eeq

\beq
\bPi_3:  \begin{tabular}{c c}
Purple & Orange \\
14 & 6 \\
14 & 6 \\
13 & 7 \\
11 & 9 \\
8 & 12
\end{tabular}
\eeq

\beq
\bPi_4: \begin{tabular}{c c}
Purple & Orange \\
12 & 8 \\
12 & 8 \\
12 & 8 \\
12 & 8 \\
12 & 8
\end{tabular}
\eeq

Each of these districtings obeys the constraint that districts be of equal size.  But the legislatures they produce are very different.  In $\bPi_1$, the Purple Party holds three seats and the Orange Party two.  In $\bPi_2$, the Orange Party holds a legislative majority, with three out of the five seats.  In $\bPi_3$, Purple holds a 4-1 majority of seats.  And in $\bPi_4$, Purple holds all five seats and Orange is utterly shut out.

Which of these choices is the most fair?  Which is the least?

With this toy case in mind, let's talk about the main existing flavors of quantitative measures of fairness.

\subsection{Proportional representation}  One of the most broadly popular and intuitively appealing measures of districting fairness is provided by the principle of {\em proportional representation}.

\begin{defn} A districting satisfies {\em proportional representation} when the proportion of seats held by each party is equal to the proportion of votes won by that party.
\end{defn}

Of the districtings above, only $\bPi_1$ satisfies proportional representation; the Purple Party got 60 percent of the vote, and it holds 60 percent of the seats.
Achieving proportional representation is often seen as a goal, or even {\em the} goal, of districting reform.  The New York Times, in a 2018 feature story on gerrymandering, wrote of Pennsylvania's Congressional districts: ``Republicans got 54 percent of U.S. House votes statewide, but won 13 of 18 seats," suggesting that this deviation of seat proportion from vote proportion is the problem districting reform is meant to solve.  Supreme Court Justice Neal Gorsuch, in the oral arguments on the redistricting case {\em Rucho v. Common Cause}, also took this to be the standard at issue, pointedly asking, ``[A]ren't we just back in the business of deciding what degree of tolerance weÕre willing to put up with from proportional representation?"

I hope it is clear at the very outset that this definition suffers from many practical problems.  For one thing, it is impossible to satisfy exactly; the proportion of votes cast for a party need not be anywhere near a rational number whose denominator is the number of districts!   This is most notable in states consisting of a single Congressional district;  we accept, without hesitation, that whoever gets the most votes should occupy 100\% of the seats, even though the proportion of votes that candidate received may be far from $1$.

Would fairly drawn maps even be likely to yield proportional representation?  It's unlikely.  Look at the Wyoming State Senate, for instance.  Wyoming is by some measures the most strongly Republican state in America.  Two-thirds of its voters picked Donald Trump in 2016, and the same proportion voted Republican in the governor's race in 2018.  But the state senate isn't two-thirds Republican; there are 27 GOP senators and only 3 Democrats.  That shouldn't necessarily be seen as unfair!  When a state's population is two-thirds Republican, the likelihood is that most geographic segments of the state are pretty Republican.  In the extreme case of this, where the state is utterly homogeneous politically, {\em every} district would be represented by a Republican Senator.  This is the situation depicted in $\bPi_4$.  By the central limit theorem, these are the kinds of districts we'd get if we selected the districting entirely at random from the set of all possible partitions of the state's population into equal-cardinality pieces, with no attention paid to geography.  Real-life states, even Wyoming, are {\em not} exactly homogeneous; but they also often don't look like $\bPi_1$, in which there's not a single district that approximates the overall political distribution of the state.  When districts are drawn geographically, an intermediate scenario like $\bPi_3$ is more common:  substantial variation from the statewide average, but with some concentration around that average.

The final problem with asking districtings to approximate a proportional representation system is that it's not {\em our} system.  We have chosen to accept, for instance, that parties with small but geographically dispersed support don't get representation in the legislature.  The proportion of Americans voting for Libertarian candidates for the House of Representatives consistently hovers around 1\%; but there has never been a representative from that party, let alone the $3-5$ that strict proportional representation would recommend.   (In Canada, whose elections are very similar to those in the U.S., the deviations are even starker; in the 2019 federal elections there, the New Democratic Party drew 16\% of the vote against only 8\% for the Bloc Qu\'{e}b\'{e}cois, but the Bloc, whose voters are concentrated in a single province, won substantially more seats in Parliament.)

This is not a matter of mathematical or purely philosophical fairness; it's a decision the United States made a long time ago, baked into the way our legal system views elections.  No matter how many party-line voters there are in practice, our votes are formally for people, not parties.

\subsection{Partisan symmetry}

\label{ss:symmetry}

One visually effective way to think about measures of fairness of a districting $\bPi$ is the {\em seats-votes curve}.  This is just what it says on the box; the locus $\set{(x,y)} \in [0,1] \times [0,1]$ consisting of points where $x$ is the proportion of overall votes going to a party and $y$ is the proportion of legislative seats that party wins.  There is a separate seats-votes curve for each party; in the present US-centric discussion, we are going to stick to cases where only two parties compete (sorry, Libertarians!) in which case the seats-votes curve for one party is the image of the seats-votes curve for the other by the transformation $(x,y) \mapsto (1-x,1-y)$.

Proportional representation is the requirement that the curve is just the line $x=y$. But here we should be careful. There are only finitely many elections held under a given district map, and so this curve is something we have only finitely many points on; what's more, it's certainly possible for two different elections to yield the same vote share but different seat shares, depending on the distribution of votes; what's {\em still} more, elections held in different years may be held under the same geographic district maps but don't represent {\em exactly} the same districting, some voters inevitably having moved out of the district, into the district, off this mortal coil, etc.  So the seats-votes curve is probably best thought of as a cloud of points around an ideal curve, and we may use as a criterion of fairness that the ideal curve has certain properties.  That it be $x=y$ on the nose is, we have argued, too much to ask.

Partisan symmetry is a much more modest request:

\begin{defn} A districting satisfies {\em partisan symmetry} if the seats-votes curve is invariant under the symmetry $(x,y) \mapsto (1-x,1-y)$.
\end{defn}

This criterion seems very natural: if the Purple Party gets 4 seats with 60\% of the vote, then the Orange Party should get 4 seats if {\em it} gets 60\% of the vote.    In particular, under the partisan symmetry constraint, a party that gets exactly half the votes gets exactly half the seats. 

One challenge for this notion is that it asks us to test whether a curve satisfies a symmetry criterion when we have access only to a set of points on the curve (or, really, a set of points {\em near} the curve.)  If we want to test whether the curve is $x=y$, that's no problem; we can use the difference between the measured $x$ and the measured $y$ as our measure on unfairness.  In order to test symmetry, we would have to probe the seats-votes curve further, inferring something about the results of elections that might have happened, but didn't.  As a simple rule of thumb, for instance, we might imagine that partisan swings are roughly uniform across districts.  So in $\bPi_1$, if we moved the overall voteshare to $50-50$, we would similarly shift $2$ votes from Purple to Orange in each of the five districts; then Purple still wins three seats, by the narrower margin of $13$ to $7$, and Orange still wins two, now in $14-6$ and $15-5$ blowouts.  This means partisan symmetry has been violated, since Purple still holds a majority of seats while getting only half the votes.  More generally, the approximate seat curve for $\bPi_1$ would be given by the step function
\beq
y = 3 \sigma(x+0.15) + \sigma(x-0.25) + \sigma(x-0.2)
\eeq
where $\sigma(x)$ is $1$ for $x > 0.5$ and $0$ for $x \leq 0.5$.  The reader can check that this seats-votes curve is not symmetric; indeed, its image under $(x,y) \mapsto (1-x,1-y)$ is the seats-votes curve for $\bPi_2$.  The only one of the four districtings we showed which satisfies partisan symmetry in this sense is $\bPi_4$.

Another criticism: partisan symmetry may reflect factors other than self-interested malfeasance.  The districting $\bPi_2$ awards a majority of seats to the Orange Party, even as they get thumped by the Purples in the popular vote.  But what if the Purples of the state are packed into a couple of dark-Purple metro areas, set against the background of a countryside that leans orange?  Isn't it possible you'd see results a lot like this, without any self-dealing?  Is ``organic partisan asymmetry" like this actually unfair?  If we ask the state to vote on a ballot referendum, people who feel strongly about the issue don't get more votes than people who barely care.  Some would apply the same reasoning to geographic regions: each patch of land gets one vote in the legislature, even though some patches may be strong supporters of one party while others are more ambivalent.

\subsection{Efficiency gap}

In the last decade, law professor Nicholas Stephanopoulos and political scientist Eric McGhee introduced and popularized a new metric for unfairness, called the {\em efficiency gap}. \cite{efficiencygap}  To see what motivates their definition, look back at our four example districtings.  What makes $\bPi_2$ such a good choice for Orange?  It's that Orange voters are deployed with exquisite strategic precision, to exactly the districts where they're needed to ensure a narrow victory. Purple voters, by contrast, are in exactly the {\em wrong} places; almost half of them reside in the three districts where Purple loses narrowly, and thus they contribute nothing to Purple's representation in the statehouse.  One might say their votes were wasted.  This leads us to a definition.

\begin{defn} Suppose the candidates in a two-party election receive $A$ and $B$ votes, respectively, with $A \geq B$. Then the number of wasted votes for the losing candidate is $B$, and the number of wasted votes for the winning candidate is $A - (1/2)(A+B)$.\footnote{We are not going to worry in this space about the difference between half the votes and half the votes plus one.}
\end{defn}

This captures the notion that a vote is wasted just insofar as it fails to contribute to a candidate's victory.  A districting drawn to favor one party does so by causing the other party to waste votes.  That motivates the definition of efficiency gap.

\begin{defn} Suppose the total number of wasted votes for the two parties in an election is $w_1$ and $w_2$ respectively, and the total number of votes cast in the election is $N$.  Then the {\em efficiency gap} is $(1/N)(w_1 - w_2)$.
\end{defn}

For example, in $\bPi_1$, the Purple party wastes $5$ votes in each of the three seats they win, and $7$ and $8$ votes respectively in the two seats they lose, for a total of $30$.  The Orange party, by contrast, also wastes $5$ votes in the first three districts, but only $3$ and $2$ in the two districts where they win, totaling to $20$.  So the efficiency gap here is $10/100 = 0.1$, favoring the Orange party.  In $\bPi_2$, the efficiency gap is much larger, $0.3$ favoring Orange.  In $\bPi_3$, the efficiency gap is $0.1$ in favor of Purple, and in $\bPi_4$ the efficiency gap is $0.3$ in Purple's direction.  This measure rates $\bPi_1$ and $\bPi_3$ as the fairest districtings; so do most people who look at those four tables, which is a point in efficiency gap's favor. 

Another bonus of efficiency gap is that, unlike partisan symmetry, it doesn't require any imputation of election results under conditions other than the real ones; it computes its measure directly from the election results that have already happened.  Efficiency gap has this in common with proportional representation, and indeed the two measures are very similar in spirit; both specify exactly what they want the seats-votes curves to be.

\begin{prop} A two-party election (with all districts of equal size) in which one party receives proportion $x$ of the votes and proportion $y$ of the seats has efficiency gap $2(x-1/2) - (y-1/2)$.  In particular, the efficiency cap is zero when $y = 2x-1/2$.
\label{egformula}
\end{prop}

The latter statement means efficiency gap can be thought of as measuring adherence to the seats-votes curve $y=2x-1/2$, instead of the curve $y=x$ required by proportional representation.  In other words, those two criteria are not only different, they are incompatible!

\begin{proof} For each district $\pi_i$, let $x_i$ be the proportion of votes won by the first party.  Let $y_i$ be $1/2$ if the first party wins and $-1/2$ if the second party wins.  So the average of $y_i$ over all districts is $y-1/2$, and the average of $x_i$ over all districts is $x$.  The proportion of votes in $\pi_i$ which are wasted votes for the first party is then $x_i - (1/2)(y_i+1/2)$, and the proportion of votes which are wasted votes for the second party is $(1-x_i) - (1/2)(-y_i+1/2)$, so the difference -- the quantity whose average over all districts is the efficiency gap -- is $2x_i - y_i - 1$.  Thus the efficiency gap is $2x-1 - (y-1/2)=2x-y-1/2$, as claimed.
\end{proof}

The efficiency gap was a huge step forward for attempts to bring mathematical reasoning to bear on the legal problem of districting.  Prior to the efficiency gap, courts had typically declined to intervene in partisan gerrymandering cases, following the Supreme Court ruling in {\em Vieth v. Jubelirer} that there was no sufficiently clear standard for distinguishing a gerrymandered map from a fair one.  The efficiency gap filled that, well, gap.  It is easy to compute, it doesn't rely on any hypotheticals, it's clearly distinct from proportional representation, and it conforms well with our native intuition about what makes a districting unfair.  It formed the centerpiece of {\em Whitford v. Gill},  the case against the state legislative districts drawn in Wisconsin after the 2010 census.  The new maps created a sharp rise in efficiency gap which persisted in several consecutive elections following redistricting.  That evidence (together with ample documentation that the map was drawn with the specific intent to help the Republican party) convinced a three-judge federal panel to throw out those districts.

But the efficiency gap isn't the end of the story.  For one thing, it is surely too strict.  Like the proportional representation standard, it asks for adherence to a specific seats-votes curve.  Do we really think one curve suits all states at all times?  The curve $2(x-1/2) = (y-1/2)$ also has the problem that it's literally impossible for the outcome to match that curve if one party gets more than $75\%$ of the vote (though admittedly this scenario is extremely rare, even in states like Wyoming where one party is much more popular than the other.)  All these difficulties are, of course, superable.  Legal arguments based on the efficiency gap don't propose a strict standard where a map with efficiency gap exceeding some threshhold are automatically declared unconstitutional; rather, a high value of the efficiency gap is to be used as a ``red flag" providing evidence, but not dispositive evidence, that a map has been gerrymandered.

Another vulnerability of the efficiency gap standard arises directly from one of its strengths.  The efficiency gap depends on real outcomes, not hypotheticals.  But that makes the efficiency gap {\em discontinuous}.  At its heart is the function $y_i$ in Proposition~\ref{egformula}, which can jump from $-1/2$ to $1/2$ with a tiny change in vote share.  A districting that fails an efficiency gap test might easily pass if a few close races switched their outcome.  That feels wrong.  (Though of course there are workarounds; one might, for instance, report the distribution of efficiency gap on a small ball around the actual outcome rather than relying on a single point)

The efficiency gap measure has another problem in the U.S. context.  American courts have generally not been sympathetic to the idea that political parties with substantial support have a constitutionally guaranteed right to legislative representation.  Claims made by individual voters that they've suffered harm to their ability to vote, or their first amendment right to express their politics, have been more successful.  So a suit filed against a district map has to argue that individual voters have had their rights removed, or at least meaningfully shaved down.  Which voters are these?  It can't just be someone whose vote was ``wasted," in the efficiency gap sense; after all, in every single district, half the votes cast are wasted, whether gerrymandering takes place or not!  The design of the efficiency gap measure is purely global; it doesn't tell you, because it wasn't designed to tell you, which districts are the ones that were maliciously modified to help one party.  A detailed critique of the efficiency gap from a mathematical perspective can be found in \cite{bernsteinduchin:eg}; for further description and several more refined measurements in the spirit of the efficiency gap, see \cite{tapp}.

\subsection{Ensemble sampling}

I have discussed a lot of simple answers to a complex question, each of which has real merits, each of which is in some ways lacking.  Now we turn to the part where deeper mathematics comes into play; I want to discuss the method of {\em ensemble sampling}, which in my view is the current state of the art for measuring gerrymandering.

Let's return to our basic question.  What does a district map that {\em isn't} gerrymandered look like?  Does it look more like $\bPi_1,  \bPi_2,\bPi_3,$ or $\bPi_4$?  We have seen that attempts to assign a numerical unfairness score to a map based on a single election outcome all have problems, and not all agree about the relative fairness of the four districtings in our toy example.

And why {\em should} they agree?  After all, fairness isn't purely a matter of the numbers on the spreadsheet; some actual knowledge of the political landscape to be partitioned is required.    If partisanship in a state is homogeneously distributed, the same in the east as in the west, in the north and in the south, then any geographically based map is going to look like $\bPi_4$, with all seats having roughly the same partisan distribution as the whole state.  In that case, one party will hold all or almost all the seats, even if their statewide share of the vote is only a modest minority.  In a state like that, a seat distribution closer to proportional representation, or to zero efficiency gap, would be strong evidence {\em for} gerrymandering, not for its absence.  Likewise, in a state where one party's support was concentrated in a geographic region, a districting like $\bPi_4$ would be ironclad proof someone had their thumb pressed firmly on the scale.

The philosophy behind ensemble sampling is a simple one:  the opposite of gerrymandering isn't proportional representation or adherence to any other strict numerical standard; the opposite of gerrymandering is {\em not gerrymandering}.  If we want to know whether a map is fair, the right question is

\begin{quote}
Does this district map tend to produce outcomes similar to a map that {\em would have been drawn} by an authority who wasn't aiming to privilege one party's interests over another?
\end{quote}

That neatly solves some of the problems with the measures described above, but at the expense of introducing a new problem, which is now not really legal or philosophical but inferential: how can we assess what would have happened if the maps had been drawn without prejudice?   The idea of studying gerrymandering through this lens was first popularized in an influential 2013 paper by the political scientists Jowei Chen and Jonathan Rodden~\cite{chenrodden}.   They were troubled by the issues above, especially the phenomenon mentioned at the end of section~\ref{ss:symmetry}:  when Democrats are predominant in cities and present in moderate numbers throughout the state, while Republicans are concentrated in more rural districts and almost entirely absent from more densely populated areas, partisan symmetry can fail even when districts are drawn indifferently to partisan advantage.  How do we distinguish an asymmetric districting like $\bPi_2$ that arises from gerrymandering from one that reflects the disinclination of Oranges to live in Purpleopolis?\footnote{Proponents of proportional representation are right to pipe up here and ask, why do we need to make this distinction?  Isn't partisan bias unfair whether or not it's enacted on purpose?  Maybe so -- but the legal and political barriers to moving away from one-geographic-district-one-seat are a lot higher than anything else described here.} 

Chen and Rodden write:

\begin{quote}
To what extent is observed pro-Republican electoral bias a function of human geography rather than intentional gerrymandering? To what extent might pro-Republican bias persist in the absence of partisan and racial gerrymandering?

The main contribution of this paper is to answer these questions by generating a large number of hypothetical alternative districting plans that are blind as to party and race, relying only on criteria of geographic contiguity and compactness. We achieve this through a series of automated districting simulations. The simulation results provide a useful benchmark against which to contrast observed districting plans.
\end{quote}

Where do the ``hypothetical alternative districting plans" come from?  That's where ensemble sampling comes in.  First of all, let's rephrase the basic fairness question along the lines of the Chen-Rodden approach:

\begin{quote}
Does this district map tend to produce outcomes similar to a map {\em randomly selected} from the set of all possible maps?
\end{quote}

This suits our intuition; one might imagine, as a first approximation, that a map-drawer indifferent to who wins would consider any partition of the population into equally-sized geographic chunks to be as good as any other.  Or one might very reasonably {\em not} assume that.  Each state has its own constitutional and other legal constraints on what districts can look like -- for example, in most states they need to be connected --  and overlaid on these is the federal Voting Rights Act, which guarantees that among the Congressional districts there are, where possible, majority-minority ``districts of opportunity."

It gets more complicated still.  Mathematicians often think of the law as consisting of a series of hard and fast rules, like axioms, from which outcomes can be drawn.  Law isn't really like that.  Law probably {\em couldn't} be like that.  In the case of districting, one quickly finds that much of the relevant law is not so much a set of constraints as a collection of preferences.  In Wisconsin, for instance, legislative districts are not supposed to cross county lines, except that decades of precedent have established that they {\em can} cross county lines when other legal requirements make it necessary, but it's better not to do it too much.  We also have a constitutional requirement that state legislative districts be ``as compact as practicable."  What does that mean?  It is certainly not a strict numerical constraint on the pair (perimeter, area) in $\R^2$.  We can capture all this by refining our question one more time:

\begin{quote}
Does this district map tend to produce outcomes similar to a map {\em randomly selected} from the set of all legally permissible maps, on which we place a probability distribution that reflects this state's legal preferences between districts?
\end{quote}
 
The standard is then that unfair maps are ones which are {\em extreme outliers} in that distribution -- that is, those that look like this:

\includegraphics{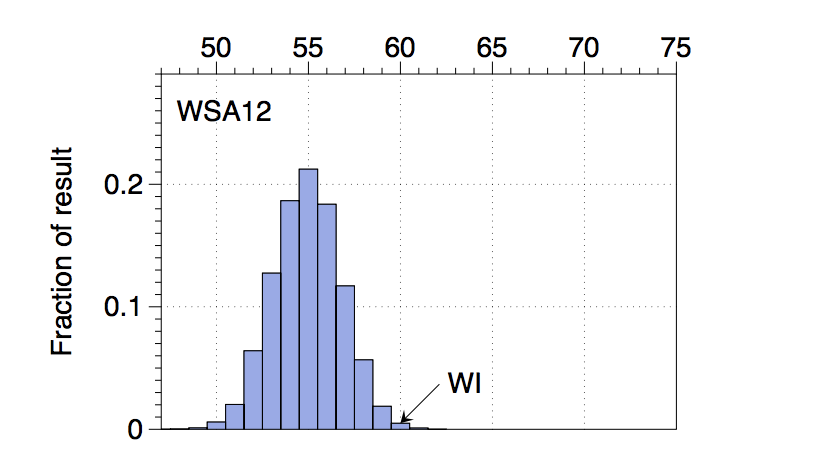}

This diagram, from a 2017 paper of Herschlag, Ravier, and Mattingly~\cite{mattinglywi}, depicts the results of a simulation of the 2012 Wisconsin State Assembly election, under 19,184 alternative districting plans sampled from their ensemble.  The outcomes form a reasonably normalish distribution centered on a modal outcome of 55 Republican seats out of 99.  In our world, under the maps drawn by the Republican majority in the state legislature.after the 2010 census, the Republicans won 60 seats.   We note in passing that Herschlag-Ravier-Mattingly's finding matches that of Chen and Rodden concerning ``unintentional gerrymandering"; Wisconsin is a state where two large urbanized areas (Madison and Milwaukee) strongly favor Democrats and only one (Waukesha County) strongly favors Republicans, and indeed, in the 2012 election where the popular vote for Assembly candidates was very close to evenly split between Democrats and Republicans, a typical map drawn without prejudice gives Republicans a modest majority of seats, though much less than the majority supplied by the gerrymandered map in actual use.

As you've probably noticed, there's a major methodological issue we've been keeping silent about.  {\em How} do we sample from the set of all permissible maps?  This is no small question, and it's the hardest mathematical part of the problem.  Wisconsin has $6,672$ voting wards.  The number of ways to partition those wards into $99$ Assembly districts is big -- really big, massively uncomputably big.  It is $99^{6672}$, if you ignore all constraints on what makes an assembly map permissible.

Now just because a set is big doesn't mean you can't uniformly sample from it.  It's easy to choose an integer uniformly at random from the interval $[0,10^{100}]$.  Or to choose a spanning tree from a graph on $1000$ vertices; you can do this efficiently via Wilson's algorithm.  Closer to the problem at hand,  the set of {\em all} partitions of Wisconsin's wards very easy to sample uniformly -- just hand each ward an integer uniformly and independently chosen from $\set{1,\ldots, 99}$.

But that's not the sample we want.  First of all, we need the districts to be of approximately equal size.  But more than that, we need the districts to be {\em contiguous}, not built out of wards scattered all over the state from Turtle Lake to Oconomowoc.  We can think of the wards as forming a {\em weighted planar graph}, where each vertices is a ward and two wards are connected if they border each other.  Then a collection of wards forms a contiguous district just when the set of vertices corresponding to that collection induces a connected subgraph.  If $\Gamma$ is a graph, we define a {\em connected $k$-partition} to be a partition of the vertices into subsets $V_1, \ldots, V_k$ such that the induced subgraph on each $V_i$ is connected.

Now we're faced with a question in the theory of algorithms:  given a weighted planar graph on $N$ nodes, how do you randomly sample from the set of connected $k$-partitions whose constituents have roughly equal total weight?  This is not so easy.  Even when $k=2$, this problem is NP-hard~\cite{najtsolomon}.

But lots of NP-hard problems have tractable approximations.  If we want to sample from an unknown distribution on the vertices of a very large graph, one can often do very well in practice by means of a random walk on the graph.  (In pure math, a well-known example of this technique is the {\em product replacement algorithm} for choosing a uniform random element of a large finite group; see \cite{gamburdpak}.)   The idea of using this kind of Markov chain in the context of redistricting originated with Mattingly-Vaughn~\cite{mattinglyvaughn} and Fifield et al~\cite{fifield}.  It has become the dominant method of ensemble sampling among mathematicians working in this area.

The graph we're sampling from is not the graph $\Gamma$ of wards described above, but a vastly huger one:  we consider a graph $P$ whose vertices are all connected $k$-partitions of $\Gamma$ -- or, better, all connected $k$-partitions corresponding to districtings compliant with state and federal law.  I've told you the vertices; what are the edges?  Here's where things get really interesting.  There are different choices, and the corresponding random walks may certainly have different behavior.  The simplest and in some ways most natural case is that of the ``flip graph," in which two vertices are adjacent in $P$ just when the corresponding partitions differ with respect to only one vertex of $\Gamma$.  That is a gigantic graph whose structure we know next to nothing about.  But given a vertex $\bPi$ of this graph, we {\em can} generate a list of its neighbors; each vertex adjacent to $\bPi$ is obtained by a ``flip":  take a ward in district $i$ which lies on the boundary with district $j$, and reassign it to district $j$, as long as this violates no legal constraint.  It's easy to list and sample from all possible flips; so we can efficiently carry out a random walk on $P$, and thereby generate a large population of legally acceptable districtings.  If you want to bias your walk towards districtings whose districts have more compact shapes, or shatter fewer counties, or whatever, you can weight your choice of moves in the random walk to promote those virtues, in the style of the Metropolis-Hastings algorithm.  In other words: the distribution we want to sample from {\em isn't} the uniform distribution, because we prefer some maps to others, so we set up our walk to get the stationary distribution we desire.  This is the walk we see used in the work of Herschlag-Ravier-Mattingly~\cite{mattinglywi} on the Wisconsin Assembly districts.  We can then identify the problematic maps as those whose electoral outcomes are extreme outliers in the sample distribution.  

Another graph structure you can put on $P$ is the {\em recombination} graph~\cite{duchin:recom}.  Suppose $\bPi$ is a districting.  Then a neighbor of $\bPi$ is obtained as follows.  Choose two adjacent districts.  Merge them into one.  Then split the new double-sized district into two connected pieces of roughly equal size.  That's a ReCom move, and two districtings are adjacent if a ReCom move sends one to the other. 

But didn't I {\em just say} the problem of sampling uniformly from the set of $2$-partitions is hard?  And isn't that just what choosing a ReCom move requires us to do?  Well, not quite.  In the ReCom process, the splitting of the double-sized district $D$ is {\em not} uniform over the set of all $2$-partitions.  Instead, we proceed as follows.  Choose a spanning tree $T$ uniformly from the set of all spanning trees of $D$; {\em this} we can do uniformly in polynomial time, by Wilson's algorithm referenced above.\footnote{The Kruskal-Karger algorithm is just as good in practice, and is sometimes used in ReCom implementations.}  Now choose an edge at random, subject to the constraint that deleting the edge splits the vertices of $T$ into two roughly equal parts $T_1, T_2$; then the vertices of $T_1$ and $T_2$ constitute a partition of the vertices of $D$ into two subsets.  Now the restriction of $P$ to the vertices of $T_i$ for each $i$ contains a tree, and so is connected; we have constructed the desired $2$-partition of $D$.\footnote{There are variants of ReCom which choose the $2$-partition in a different way, but for simplicity we'll stick to spanning tree recombination in these notes.}

\begin{figure*}
\caption*{A step in the ReCom process (Figure 4 in \cite{duchin:recom})}
\includegraphics[scale=0.3]{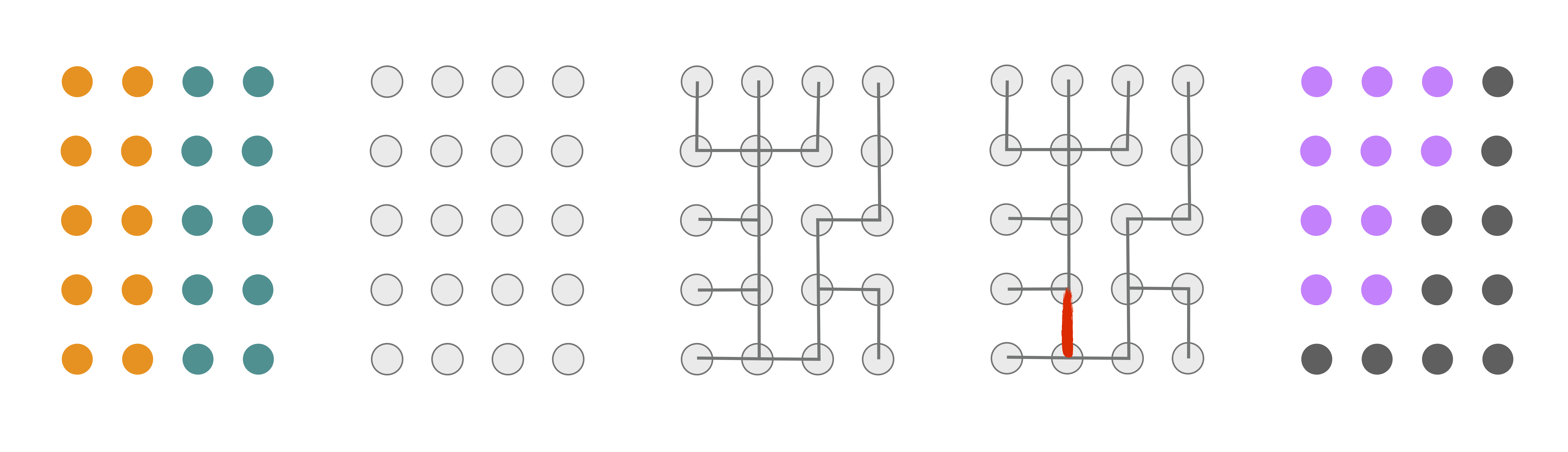}
\end{figure*}
(I can't resist pointing out in passing that the set of spanning trees of a planar graph has an extremely interesting structure.  By Kirchhoff's matrix-tree theorem, the number of spanning trees is the same as the order of the finite abelian group alternately called the sandpile group, the $\Pic^0$ of the graph, or the tropical Jacobian.  When the graph is planar, a beautiful categorification of this numerical identity holds:  the set of spanning trees is canonically a {\em torsor} for the sandpile group.!\cite{chan},\cite{baker}.  Is there any chance that the tropical viewpoint on graphs is useful for computation?)

The non-uniformity of the choice of the $2$-partition might at first seem artificial, but in fact it's more feature than bug!  A uniformly chosen $2$-partition typically has a long, snaky boundary, a sort of space-filling curve.  The partitions coming from the spanning tree method, by contrast, are strongly biased towards having short boundaries; in graph-theoretic terms, there tend to be few edges of $D$ joining vertices of $T_1$ to vertices of $T_2$.  For the flip walk, we need to build a preference for compact districts into the weights of the random walk if we want to get decent-looking districts; ReCom, by contrast, tends to form compact districts without any extra infrastructure.  What's more, the ReCom walk appears to converge to a stationary distribution more efficiently than does the flip walk, although it is more difficult to describe in explicit terms what the stationary distribution is.  Gaining a better understanding of the stationary distributions attached to various random walks on $k$-partitions is one of the richest open problems in the subject, and is of interest not only in practical terms but as a question in pure stochastic processes.

Yet another appealing feature of the ReCom walk is that software to do it is open-source and publicly available~\cite{gerrychain}; I encourage you to mess with it yourself!

Ensemble sampling has proven to provide a more effective, convincing, and intuitive quantitative measure of gerrymandering than those before it.  It was the centerpiece of the gerrymandering cases presented to the Supreme Court in 2019; a bipartisan twin set, one ({\em Rucho v. Common Cause}) addressing a district map gerrymandered by Republicans in North Carolina, the other ({\em Lamone v. Benisek}) a map gerrymandered by Democrats in Maryland.  I don't know if this is the most mathematical case the Supreme Court has ever addressed, but I believe it is the first time the Court  received a "Mathematicians' Brief," an amicus brief signed by eleven mathematical scientists, including me, explaining the quantitative aspects of the case.

In {\em Rucho} and {\em Lamone}, I hurry to point out, the Supreme Court ruled that partisan gerrymandering was not justiciable; that is, it was not a matter where the federal courts had the right to intervene.  To anyone who had been following the mathematical study of redistricting, the court's decision was puzzling.  Justice Roberts, writing for the majority, says:

\begin{quote}
Partisan gerrymandering claims invariably sound in\footnote{As best I can tell from my lawyer friends, ``sound in" here means something in between ``derives from" and "amounts to" -- and people say {\em we} talk in impenetrable jargon!} a desire for proportional representation. As Justice O'Connor put it, such claims are based on Òa conviction that the greater the departure from proportionality, the more suspect an apportionment plan becomes.Ó 
\end{quote}

I am not weighing in on the legal merits of the decision when I say that this claim is wildly off the mark.  As we have seen, proportional representation is not the aim of any modern measure of electoral fairness, or any claim brought before the court in the 2019 districting cases  (It may indeed have been an issue 33 years ago, when Justice O'Connor wrote the words Roberts quotes in her concurrence in {\em Davis v. Bandemer})  Indeed, one measure arising in contemporary cases, the efficency gap, is {\em incompatible} with proportional representation, while the ensemble sampling methods (correctly, in my view) are orthogonal to proportional representation.  In some case a map that yields proportional-representation outcomes would be licensed by those methods; in other cases, like U.S. congressional districts in Massachusetts, a map yielding proportional representation would be flagged as an extreme outlier~\cite{duchin:mass}.

To be clear, the Supreme Court did not take seriously any claims that gerrymandering had not taken place, and indeed the majority decision endorsed the view that the practice of gerrymandering is ``incompatible with democratic principles."  The Court's decision rests on the fact that some things are incompatible with democratic principles but not incompatible with the Constitution.  When this is held to be the case, the federal judiciary walks on by with eyes modestly averted.  But the Court did portray partisan gerrymandering as a problem somebody -- just not the Justices themselves -- ought to remedy.  

And indeed, the arguments derived from ensemble sampling have found more purchase elsewhere.  In 2018, the Pennsylvania Supreme Court threw out the state's U.S. Congressional district map, relying heavily on expert testimony from Jowei Chen and from Wesley Pegden, a mathematician at Carnegie Mellon (more on Pegden's work just down the page!) identifying the map as an extreme partisan outlier from a random walk ensemble.  Pegden was subsequently appointed by Governor Tom Wolf to serve on the Pennsylvania Redistricting Reform Commission, along with the more traditional group of elected officials and community leaders you might expect to find on a political panel like this.  Wolf also enlisted Moon Duchin to help evaluate replacement districting plans.  In North Carolina, a state court similarly found that state legislative districts drawn by the majority in the state legislature violated both the Free Elections Clause and the Equal Protection Clause of the North Carolina Constitution.  Their decision, like the one in Pennsylvania, is deeply rooted in the analysis of ensemble samples, citing testimony of Jonathan Mattingly as well as Pegden and Chen.  (To get a sense of what kind of testimony mathematicians provide, you could look at Mattingly's expert report in that case~\cite{mattingly:expertreportnc}.)  One key point from a legal perspective is that ensemble methods can make local assessments, not just global ones as the efficiency gap does, identifying {\em individual districts} as outliers, which helps courts in several ways.  It makes it easier identify harm to individual voters; it can provide courts with a remedy that throws out only part of the map instead of the whole thing; and, most importantly, it presents judges with a clearer picture of {\em how} gerrymandering is accomplishing its goals.

I don't want to leave the impression that ensemble sampling is an infallible gerrymandering detector.  Many challenges, some purely mathematical, others intertwining math and law, remain.

A central open question concerns speed of mixing.  A random walk on a connected graph converges to the stationary distribution.  But how {\em quickly} it converges is a delicate matter, involving, among other things, the spectral gap of the adjacency matrix of the graph.  For the graph $P$ of $k$-partitions, we have no control over mixing time, and so no guarantee that the ensembles obtained by random walk are drawn from the stationary distribution.  Indeed, if we impose on $P$ the condition that the $k$ components of the partition have roughly the same size, as far as I know there is no proof that $P$ is even connected!  Even if $P$ is connected, it might have a ``long neck" connecting two large regions

If our random walk starts on one of those regions,, crossing the neck is a very low-probability event; so even if we run the random walk for a long time, we may be approximating a distribution which is far from the stationary one, but rather approximates a distribution supported on one side of the neck.   

From a political and legal perspective, it's not clear this matters.  Suppose a district map $\bPi$ tends to give the Purple Party $60$ out of $100$ seats, and suppose this figure is an outlier in a sample of maps near $\bPi$ in $P$, $99.9\%$ of which give the Purple Party at most $55$ seats.  One cannot strictly rule out the existence of an ``undiscovered country" of districtings in which $60$ Purple seats are the norm.  But the ensemble sample still feels like very strong evidence that this map $\bPi$, so unusual among its neighbors, was not picked out indifferently to its Purple-friendliness.  This insight was formalized and made into a theorem by Chikina, Frieze, and Pegden (\cite{chikina}, see also \cite{chikina2}) who obtain a provable threshold for statistical significance without any need for a bound on mixing time.   We'll state their result in the language of random walks on graphs, though it actually holds for any reversible Markov chain.\footnote{The flip walk, in its most commonly used form, is reversible, but the ReCom walk isn't; see \cite{mergesplit} for a reversible modification of ReCom.} They prove:  for any small $\epsilon > 0$, and any real-valued score function $\omega$ on the vertices of a graph $G$, the probability that a vertex $v_0$ of $P$ chosen uniformly (the null hypothesis) has a higher score than $1-\epsilon$ of the first $k$ vertices in a random walk starting from $v_0$ is at most $\sqrt{2\epsilon}$.  This allows you, if you like, to assess gerrymandering using traditional frequentist-statistical tools like statistical significance.

Nonetheless, the question of what kind of mixing time one expects for random walks of various flavors on sets of $k$-partitions of planar graphs is a really interesting one.  In practice, as we have mentioned, the ReCon walk seems to converge substantiallly more efficiently than does the flip walk.  Why?  And are there still more effiicient graph structures on $P$ out there to be exploited?  The study of mixing on these graphs has the flavor of statistical physics and the theory of self-avoiding random walks; see the recent paper of Najt, DeFord, and Solomon~\cite{najtsolomon} for this connection, together with questions about the way different discretizations of the same planar landscape can yield surprisingly large differences in the behavior of the corresponding random walks.

Another challenge of ensemble sampling is that different elections are different.  One certainly doesn't want to say that an individual voter is, once and for all, a Democrat or a Republican whose voting behavior is independent of time and the individual candidates on the ballot.  If we did say this, empirical data would contradict us.  In practice, what this means is that a given map may be an extreme outlier in some elections and not in others.  Take, for instance, the map of Wisconsin state assembly districts.  In 2012, a year when the statewide vote for Assembly seats in Wisconsin was very close to 50-50, the result of 60 seats for Republicans was very different from the ensemble-modal value of 55 seats.  But two years later, in 2014, the electorate leaned much more towards Republicans; and in that election, the 63 seats won by Republicans sit comfortably in the middle of the range of outcomes produced by the ensemble.

So is the Wisconsin district map an extreme outlying gerrymander, or is it not?  The work of Herschlag, Ravier, and Mattingly~\cite{mattinglywi} provides critical insight here.  A district map, remember, is made without knowledge of exactly who is going to vote, or how.  The dark art of gerrymandering has to be robust to this basic uncertainty.  And there are tradeoffs:  maximizing the extent to which the map helps your party under one set of circumstances may make the map less effective under other conditions, or even hurt your party if things go really sideways.  Herschlag-Ravier-Mattingly find that the Wisconsin map is designed as a sort of ``firewall."  In electoral environments where the statewide vote leans Republican, the gerrymander doesn't do much work.  But when the electorate is split evenly or even leaning slightly Democratic, it provides a powerful force towards maintaining a Republican majority in the assembly.  The gerrymander works exactly when, at least according to the desires of its makers, it needs to work, locking in a Republican seat majority over the whole range of statewide electoral conditions one might reasonably expect to encounter in an evenly split state like Wisconsin.

\section{What's next?}

The study of random walks applied to districting has developed very rapidly in the last five years, and provides opporunities for an extraordinarily direct interaction between advanced mathematics and public policy.  But the story is far from over, and there's a lot of work still to do for mathematicians interested in these problems.

Part of the work is outreach.  Moon Duchin and Jonathan Mattingly have both launched centers -- respectively, the Metric Geometry and Gerrymandering Group at Tufts, and the Quantifying Gerrymandering group at Duke -- which serve as clearinghouses for new research on districting and launchpads for both early-career and senior mathematical scientists interesting in getting involved.

There is also much more mathematics to do, besides the rich vein of questions indicated above about mixing times for Markov process on the k-partitions of planar graphs.  One key question, which has been raised a lot but so far has not been extensively addressed, is:  what to do about situations where there are more than two parties?  Suppose I don't care for either Purple nor Orange, preferring the Ecru Party to either one.  But perhaps, if Ecru is not an option, I like Orange better than Purple.  In a first-past-the-post system like that in the U.S., my voting behavior may depend on the district I'm drawn into.  If Ecru has no chance and Orange and Purple are close, I may vote Orange to stop Purple.  But if Purple is safely ahead in my district, I'm more likely to vote my Ecru conscience.  {\em Everything} about voting behavior is much harder to analyze when more than two parties have substantial support; the problem of districting is no exception.  In a U.S. context, where parties other than Democrats and Republicans are very rarely competitive for legislative seats, one might this question can be ignored.  But the U.S. is not the only country with geographic districts.  What's more, many American jurisdictions, including the entire state of Maine, have abandoned first-past-the-post in favor of ranked-choice voting.  To the extent RCV becomes mainstream, there will likely be more votes for candidates other than Democrats and Republicans; it seems prudent to have the mathematical machinery for analyzing districtings ready in advance.

What is the future for this interaction between mathematics and politics?  The Supreme Court has, for now, put an end to the idea that political parties nationwide will be forbidden from extreme gerrymandering.  In some states, like Pennsylvania and North Carolina, courts will throw out existing gerrymandered maps; in others, like Ohio and Michigan, legislation or ballot initiatives will delegate the process of redistricting to non-partisan panels.  In both cases, though, there are still fundamental design questions: if the process of drawing the map isn't to be  ``party operatives in a smoke-filled room tweak and twist the districting until it delivers every possible advantage to their party," what {\em is} the right process? 

One question that comes to many mathematicians' minds at this point is:  if we can generate an ensemble of thousands and thousands of potential district maps, which are compliant with the Voting Rights Act and other legal constraints, which perform well on traditional districting criteria like compactness and  county-splitting, and which are completely indifferent to which party does better, why don't we \ldots just pick one of those maps at random and call it a day?

The reasons are political.  Which doesn't mean they're not good reasons!  For one thing, algorithmic maps will inevitably miss criteria specific to the case at hand that are legitimately important to constituents.  As DeFord, Duchin, and Solomon write in their report on districting alternatives in Virginia~\cite{mggg:virginiacomp}, ``We emphasize that these ensemble methods should not be used to select a plan for enactment because they are made without local and community-based considerations. Instead, ensemble methods give an effective means of verifying whether a newly proposed plan is an extreme outlier in the universe of valid plans."  Even when used as a benchmark, a process like ensemble sampling is viewed with some suspicion by political actors.  They can't see what's under the hood.  To actually hand over map-drawing power to the algorithm is something neither elected officials nor their constituents are likely to swallow.  To a lesser extent, proposals to delegate the power to an independent nonpartisan commission -- say, retired judges or a panel of state residents -- meet the same resistance (from legislators, if not from their constituents, who have recently voted for such plans by referendum in Colorado, Michigan, Missouri, and Utah.)  Elected officials, as a rule, don't like relinquishing powers the existing framework affords them. 

But how can there possibly be a fair protocol for district-drawing if the district-drawing is to be done by the legislature itself?  In \cite{pegdengame} Pegden, Procaccia, and Yu propose a really interesting idea, deriving from an entirely different area of math:  the theory of fair-division games.  That theory descends from a single algorithm:  "I cut you choose."  Two players, who perhaps don't trust or even like each other, want to divide a cake, and each wants to make sure the other doesn't get more than their share.  (See the relevance?)  Algorithm:  one player cuts the cake in two pieces, and the other picks which piece they get.  The cutter has an incentve to make the division as close to 50-50 as possible, and the chooser, if the cutter does their job properly, is indifferent to the choice they're presented with.  The algorithm doesn't {\em enforce} fairness; but it leaves both parties to the decision feeling like they had a fair chance to affect the outcome.

The Pegden-Procaccia-Yu protocol is called ``I cut you freeze."  It's a game where Purple and Orange take turns.  They start with the district map $\bPi = \bPi_1, \ldots \bPi_k$ as it currently exists.  At each stage, some subset of the districts is {\em frozen} -- the boundaries of those districts are fixed and can no longer be modified on later turns.   Each turn has two parts:  you freeze one of the not-yet-frozen districts, then you redistrict the unfrozen part of the map however you like, then you pass the new map over to the other player.  The game ends when all the districts are frozen, each party having locked in the final form of half the districts.  (This might be better called "I freeze and cut, then you freeze and cut," but I think the authors' choice to contract this for euphony was a wise one.)  Note, crucially, that the district a player gets to freeze is one chosen from a map {\em created by the other party}; if the steps were reversed, so that each party redistricted the unfrozen part of the map first and then froze a district of their own making, it would be substantially easier for parties to create mischievous districts.

There are lots of possible game protocols one could use for redistricting, and the Pegden-Procaccia-Yu paper has inspired several competing proposals.  Understanding which, if any, provides a suitably effective buffer against extreme gerrymandering while remaining palatable to elected officials and compliant with legal requirements is a rich and fundamentally interdisciplinary question which we've just started to penetrate.

So far, the role of mathematicians in districting has been a form of damage control -- we come in to assess the outcome of decisions that have already been made, under a regime of rules  already been set in place.  The work of Pegden, Procaccia, and Yu is an example of a deeper interaction that's just starting to take shape, namely:  which regimes of rules would lead to better outcomes?  This is especially important at this time of flux, when the issue of districting is at its highest political salience in years and many states are launching brand-new commissions.  Decisions about the design of the districting process and the rules governing it are going to have consequences for decades to come, and it seems like a really good idea for mathematicians to be in the room when those decisions are being made.\footnote{Maybe we could have kept Missouri from enshrining in its state constitution the principle that  ``In general, compact districts are those which are square, rectangular, or hexagonal in shape."  It's ironic that Missouri, a trapezoid, would be so hostile to general quadrilaterals.}  Considering different rules and figuring out what would follow from them is {\em kind of our thing.}  

Ensemble sampling is well-suited for the task because it enables us to efficiently explore the space of what's possible under various collections of rules. People may once have thought that requiring districts to have equal population ensures fair representation; we now know that's not true.  Nor are traditional geometric notions like "compactness" enough.  The picture associated with ``gerrymandering" for most people is a bizarrely shaped branching snake of a district, drawn to its odd contours to achieve an electoral result.  There are districts like that in real life, but it turns out requiring districts to be roughly round~\footnote{in whatever sense: see \cite{duchintenner} for a suggestion of a discretized notion of compactness which seems better suited for modern applications than the many, many traditional methods in use.} is also not sufficient to prevent strong partisan gerrymandering.  Ensemble methods have also shown that rules which are facially neutral to party, like maximization of the number of competitive districts, can introduce partisan bias~\cite{competitiveness}.  What's more, the ensemble method speaks to much more than the number of seats each party wins -- that's just one statistic attached to the maps in the ensemble.  Which rules promote, or suppress, the power of minority voters?  Which tend to lead to more competitive districts?  What are the tradeoffs between properties of district maps we think of as virtues?  The mathematics of redistriciting isn't just a gerrymandering detector; it is, at least potentially, a full-fledged tool for the exploration of the mysterious space of protocols for representative democracy.

Whether the future of districting in the United States is independent commissions put in place by popular ballot initiatives, intricate games of bipartisan cake-cutting, or the status quo of entrenched parties grimly maximizing their own interest, is a political question.  But it's a political question shot through with mathematical content, and thanks to the work discussed here, the mathematical community has gotten engaged with this content to an extent rarely seen in American politics.  I hope this engagement continues, and I hope some of the audience here will become part of it!
 
 \bibliographystyle{amsplain}


\bibliographystyle{amsplain}

\begin{thebibliography}{99}
 
 \bibitem{bernsteinduchin:eg}{Bernstein, M. and Duchin, M., ``A formula goes to court: Partisan gerrymandering and the efficiency gap." {\em Notices of the AMS} 64.9: 1020-1024 (2017)}
 
 \bibitem{baker}{Baker, M., and Wang, Y. The Bernardi process and torsor structures on spanning trees. {\em International Mathematics Research Notices}, 16), 5120-5147. (2018)}
 
 \bibitem{mergesplit}{Carter, D., Herschlag, G., Hunter, Z., and  Mattingly, J. A Merge-Split Proposal for Reversible Monte Carlo Markov Chain Sampling of Redistricting Plans,  arXiv:1911.01503. (2019)}
 
 \bibitem{chan}{Chan, M., Church, T., and Grochow, J. A. Rotor-routing and spanning trees on planar graphs. {\em International Mathematics Research Notices}, 2015(11), 3225-3244. (2015)}

 \bibitem{chenrodden}{Chen, J., and Rodden, J.. ``Unintentional gerrymandering: Political geography and electoral bias in legislatures." Quarterly Journal of Political Science, 8(3), 239-269. (2013)}
 
 \bibitem{chikina}{Chikina, M. and Frieze A., and Pegden, W. ``Assessing significance in a Markov chain without mixing." {\em Proceedings of the National Academy of Sciences} 114.11: 2860-2864 (2017).}
 
 \bibitem{chikina2}{Chikina, M., Frieze, A., Mattingly, J., and Pegden, W. ``Practical tests for significance in Markov Chains." arXiv:1904.04052. (2019)}
 
 \bibitem{competitiveness}{DeFord, D., Duchin, M., and Solomon, J., ``A Computational Approach to Measuring Vote Elasticity and Competitiveness," (2019)}
 
 \bibitem{duchin:recom}{DeFord, D., Duchin, M., and Solomon, J., ``Recombination:  a family of Markov chains for redistricting" (2019, forthcoming)}
  
 \bibitem{duchin:expertpa}{Duchin, M., ``Outlier analysis for Pennsylvania congressional redistricting," (2018)}
 
 \bibitem{duchin:mass}{Duchin, M., Gladkova, T., Henninger-Voss, E., Klingensmith, B., Newman, H.,  and Wheelen, H. ``Locating the representational baseline: Republicans in Massachusetts" arXiv:1810.09051. (2018)}
 
 \bibitem{duchintenner}{Duchin, M., and Tenner, B. E. ``Discrete geometry for electoral geography." arXiv preprint arXiv:1808.05860. (2018)}
 
 \bibitem{fifield}{Fifield, B., Higgins, M., Imai, K., and Tarr, A. A new automated redistricting simulator using markov chain monte carlo. Work. Pap., Princeton Univ., Princeton, NJ (2015).}
 
 \bibitem{gamburdpak}{Gamburd, A. and Pak, I., ``Expansion of product replacement graphs," {\em Combinatorica} 26.4: 411-429 (2006)}
 
 \bibitem{mattinglywi}{Herschlag, G., Ravier, R., and Mattingly, J., ``Evaluating Partisan Gerrymandering in Wisconsin", arXiv: 1709.01596. (2017)}
 
 \bibitem{mattingly:expertreportnc}{Mattingly, J. ``Expert Report on the North Carolina State Legislature," available at \url{https://sites.duke.edu/quantifyinggerrymandering/files/2019/09/Report.pdf}}
 
 \bibitem{mattinglyvaughn}{Mattingly, J. C., and Vaughn, C. Redistricting and the Will of the People. arXiv preprint arXiv:1410.8796 (2014)}
 
 \bibitem{gerrychain} Metric Geometry and Gerrymandering Group. {\bf GerryChain} available at \url{https://gerrychain.readthedocs.io/en/latest/}.
 
 \bibitem{mggg:virginiacomp}{Metric Geometry and Gerrymandering Group. ``Comparison of Districting Plans for the Virginia House of Delegates," (2018)}
 
 \bibitem{najtsolomon}{Najt, Lorenzo, Daryl DeFord, and Justin Solomon. ``Complexity and Geometry of Sampling Connected Graph Partitions." arXiv: 1908.08881. (2019)}
 
 \bibitem{pegdengame}{Pegden, W., and Procaccia, A.  and Yu, D. ``A partisan districting protocol with provably nonpartisan outcomes."  arXiv:1710.08781 (2017)}
 
 \bibitem{efficiencygap}{Stephanopoulos, N.., and McGhee, E.. "Partisan gerrymandering and the efficiency gap." {\em U. Chi. L. Rev. 82}: 831 (2015)}
 
 \bibitem{tapp}{Tapp, K. Measuring Political Gerrymandering. {\em The American Mathematical Monthly}, 126(7), 593-609. (2019)}
 
 \end{thebibliography}

\end{document}